\documentclass[unsortedaddress,reprint]{revtex4-1}

\usepackage[utf8]{inputenc}
\usepackage{amssymb,latexsym}
\usepackage{amsbsy} 
\usepackage{amsmath,amsthm}

\usepackage[shortlabels]{enumitem}
\usepackage{hyperref}

\usepackage{graphicx}
\usepackage{color}

\theoremstyle{plain}

\theoremstyle{remark}

\theoremstyle{definition}

\newcommand \commentout[1] {}

\begin{document}

\title{The Coupled-Cluster Formalism -- a Mathematical Perspective}

\author{A. Laestadius 
	}
\email{andre.laestadius@kjemi.uio.no}

\author{F. M. Faulstich}

	\affiliation{
		Hylleraas Centre for Quantum Molecular Sciences, Department of Chemistry, University of
		Oslo, P.O. Box 1033 Blindern, N-0315 Oslo, Norway}

\date{\today}

\begin{abstract}
The Coupled-Cluster theory is one of the most successful high precision methods used to solve the stationary Schr\"odinger equation. 
In this article, we address the mathematical foundation of this theory with focus on the advances made in the past decade.
Rather than solely relying on spectral gap assumptions (non-degeneracy of the ground state), we highlight the importance of coercivity assumptions -- G\aa rding type inequalities -- for the local uniqueness of the  Coupled-Cluster solution. 
Based on local strong monotonicity, different \emph{sufficient} conditions for a local unique solution are suggested. 
One of the criteria assumes the relative smallness of the total cluster amplitudes (after possibly removing the single amplitudes) compared to the G\aa rding constants.
In the extended Coupled-Cluster theory the Lagrange multipliers are wave function parameters and, by means of the bivariational principle, we here derive a connection between the exact cluster amplitudes and the Lagrange multipliers.
This relation might prove useful when determining the quality of a Coupled-Cluster solution.
Furthermore, the use of an Aubin--Nitsche duality type method in different Coupled-Cluster approaches is discussed and contrasted with the bivariational principle.
\end{abstract}

\maketitle

\section{Introduction}
One of the most successful high accuracy \textit{ab initio} computational schemes is the Coupled-Cluster (CC) approach \cite{bartlett2007coupled}. 
It goes back to Coester \cite{coester1958bound}, who in 1958 suggested using an exponential parametrization of the wave function. 
This parametrization was derived independently by Hubbard \cite{hubbard1957description} and Hugenholtz \cite{hugenholtz1957perturbation} in 1957 as an alternative to summing many-body perturbation theory (MBPT) contributions order by order.
At that time, Coester was not able to come up with working equations that one might try to solve.
Those were presented by {\v{C}}{\'\i}{\v{z}}ek~\cite{vcivzek1966correlation} after the relevant concepts had been introduced in the context of quantum chemistry.
In this work, {\v{C}}{\'\i}{\v{z}}ek mentioned the projective approach of the equations, which is exploited in all conventional CC methods until today.
Firstly, in~\cite{vcivzek1966correlation} the working amplitudes and energy equations were derived when the cluster operator is approximated by merely double excitations. 
Secondly, the CC theory was compared with MBPT, configuration interaction (CI), and the pair cluster expansions of Sinanoglou \cite{sinanouglu1962many}.
Thirdly, the first ever CCD and linearized CCD computations were reported for nitrogen and a model of benzene.  
For a more detailed description of the CC history, we refer to reviews by pioneers of the theory.
For example, K\"ummel \cite{kummel1991origins} and {\v{C}}{\'\i}{\v{z}}ek \cite{vcivzek1991origins} wrote such articles within the workshop 'Coupled Cluster Theory of Electron Correlation'.
Furthermore, see the articles by Bartlett~\cite{bartlett2005theory}, Paldus~\cite{paldus2005beginnings}, 
Arponen~\cite{arponen1991independent} and Bishop~\cite{bishop1991overview}. \newline
\indent Unlike the CI method, the CC formalism does not arise from the Rayleigh--Ritz variational principle and is therefore said to be non-variational in that sense. 
This yields the well-known fact that the CC energy is in general not equal to the expectation value of the Hamiltonian and in general not an upper bound to the ground-state energy.
The reliability of quantum chemical methods is in most cases based on benchmarking, and the results' physical and chemical consistency with existing theory. 
The \textit{gold standard of quantum chemistry} -- the CCSD(T) method~\cite{raghavachari1989fifth,bartlett1990non} -- is no exception of this. 
It is the importance of sharp statements of an \textit{ab initio} method's reliability that is the motivation of this work.
Here, we build on a local analysis~\cite{Schneider2009analysis} of the CC theory that also holds in the exact, so-called \emph{continuous}, formulation with infinitely many one-particle basis functions~\cite{Rohwedder2013continuous,Rohwedder2013error}.\newline
\indent There is a rich history of mathematical investigations addressing CC methods prior to the local analyses in \cite{Schneider2009analysis,Rohwedder2013continuous,Rohwedder2013error}.
To give a complete historical account is beyond the scope of this article. We therefore limit ourselves and mention only a few 
important results. 
As a system of polynomial equations, the CC equations can have real or if the cluster operator is truncated, complex solutions.
Furthermore, using quasi-Newton--Raphson methods to compute solutions of non-linear equations can lead to divergence since the approximated Jacobian may become singular. 
This is, in particular, the case when strongly correlated systems are considered.
These and other related aspects of the CC theory have been addressed by~{\v{Z}}ivkovi{\'c} and Monkhorst~\cite{vzivkovic1977existence,vzivkovic1978analytic} and Piecuch {\it et al.}~\cite{piecuch1990coupled}.
Significant advances in the understanding of the nature of multiple solutions of single-reference CC have been made by~{\v{Z}}ivkovi{\'c} and Monkhorst~\cite{vzivkovic1978analytic}, Kowalski and Jankowski~\cite{kowalski1998towards}, and by Piecuch and Kowalski~\cite{piecuch2000computational}.
An interesting attempt to address the existence of a cluster operator and cluster expansion in the open-shell case was done by Jeziorski and Paldus~\cite{jeziorski1989valence}. 
We would also like to mention the coupled-electron pair approximation (CEPA) \cite{meyer1971ionization,meyer1973pno,ahlrichs1975pno,kutzelnigg1977methods}. This approach was introduced as a size-consistent alternative to the CISD method that was achieved by modifying (through topological factors \cite{ahlrichs1985coupled}) the CI equations to account for higher excitations. This makes CEPA non-variational (for an adapted variational formulation of CEPA see \cite{kollmar2010coupled}). CEPA can be regarded as an approximation of the CC method and does not form a truncation hierarchy that converges to the full-CI limit \cite{wennmohs2008comparative}.  \newline
\indent Mathematical analysis is a well-established part of many natural sciences. 
Plenty examples show how various fields benefit from mathematical rigor and that mathematical analysis can define a framework of the method's applicability.
This work takes off from recent developments of local analyses of CC methods, including the single-reference CC, the extended CC, the tailored CC (TCC) and its special case the CC method tailored by tensor network states (TNS-TCC) \cite{Schneider2009analysis,Rohwedder2013continuous,Rohwedder2013error,laestadius2017analysis,faulstich2018analysis}.
In the spirit of Robert Parr's fundamental approach to quantum chemistry, which was honored during the 58{\it th} Sanibel Symposium, we here present some mathematical concepts used to analyze CC methods in a functional analytic framework.
These yield rigorous analytical results that are independent of benchmarks and interpretations but rather based on mathematical assumptions. 
Adapting these assumptions to cover the computations performed in practice remains a challenge and is subject of future work.
The local analysis puts as a sufficient -- but not necessary -- condition that the cluster amplitudes are small relative to other constants.
We discuss a possible way out of this restriction motivated by the fact that CC calculations are known to work for large (single) amplitudes as well.
We furthermore address the $t_1$-diagnostic~\cite{Lee1989} and mathematically derive a more sophisticated strategy that includes all cluster amplitudes and offers a sufficient condition of a locally unique and quasi-optimal solution (after possibly rotating out the single amplitudes) rather than rejection based on just large single amplitudes. 
We furthermore complement the literature by a detailed discussion on spectral gap assumptions. 
In this context, spectrum refers to the point spectrum, i.e., the eigenvalues of relevant operators. 
Although a gap between the highest occupied molecular orbital and the lowest unoccupied molecular orbital (HOMO-LUMO gap), or a spectral gap of the exact Hamiltonian $\hat H$ (non-degenerate ground state), is crucial for the analysis, we highlight the importance of coercivity conditions, either for $\hat H$ or the Fock operator $\hat F$.
Additionally, we derive an optimal constant in the monotonicity proof of the CC function for the finite dimensional case, i.e., the projected CC theory.
Comparing the CC Lagrangian with the extended CC formulation~\cite{laestadius2017analysis}, we propose by means of the bivariational principle an alternative to measure the quality of the Lagrange multipliers, here interpreted as wave function parameters.  \newline
\indent This article is structured as follows: In Section~\ref{sec:ExpManifold}, a brief summary of the CC theory is presented. We introduce the set of admissible wave functions and moreover define cluster operators, the CC function, and the CC energy (for a full scope treatment of the mathematical formulation of CC theory presented here we refer to \cite{Rohwedder2013continuous,Rohwedder2013error}). 
In Section~\ref{sec:loc}, we discuss the use of local analysis within different CC methods. 
Key concepts here are (see Section~\ref{sec:locDefs} for definitions) {\it local strong monotonicity} and {\it local Lipschitz continuity} of the CC function $f$, which -- if fulfilled -- are sufficient conditions for a locally unique solution of $f=0$ by Zarantonello's theorem. 
In particular, the importance of so-called G\aa rding inequalities is demonstrated. 
This is done both for the Hamiltonian, Section~\ref{sec:GardingH}, and for the Fock operator, Section~\ref{sec:GardingF}. 
We conclude in Section~\ref{sec:Numer} with an overview of the Aubin-Nitsche method and the bivariational principle as they are used in CC methods for estimating the truncation error of the energy.\newline
\indent The authors are thankful to the organizers of the 58$th$ Sanibel Symposium under which many ideas presented here took form.
Moreover, the anonymous referee greatly improved a previous draft of this article -- especially putting the local analysis, under consideration here, into the context of the rich quantum chemistry literature on CC methods.
This work was supported by the European Research Council (ERC-STG-2014) through the Grant No. 639508, and furthermore supported by the Norwegian Research Council through the CoE Hylleraas Centre for Quantum Molecular Sciences Grant No. 262695. AL and FMF thank Simen Kvaal and Thomas Bondo Pedersen for useful comments and discussions.

\section{Wave Functions on an Exponential Manifold}
\label{sec:ExpManifold}
The aim of electronic many-body methods, such as the CC approach, is to solve the electronic Schr\"odinger equation (SE) $\hat H\psi = E_0\psi$ of an $N$-electron system. 
Here, $E_0$ is the ground-state energy and $\hat H$ the self-adjoint Coulomb Hamiltonian.
In this work, we restrict our attention to real Hamiltonians and wave functions.
We emphasize that the mathematical framework of Hermitian operators is not sufficient to support the necessary spectral theory for quantum mechanics. 
Thirring exemplified this with the radial momentum operator $\hat P_r = -\mathrm{i}\hbar \frac{\partial}{\partial r}$ on $\mathcal{D}(\hat P_r) = \{ \psi \in L^2((0,\infty)) : \psi(r=0)=0 \text{ and } \hat P_r\psi \in L^2((0,\infty)) \}$ \cite{schrodinger1987schrodinger}. \newline
\indent From a mathematical viewpoint the Coulomb Hamiltonian, like most differential operators, is studied in its \emph{weak form} to allow a larger variety of solutions. Set $\Omega = \mathbb R^3\times \{ \pm\frac 1 2\}$ (or any other appropriate region in space and number of spin states) and let $\int_{\Omega^N} \mathrm{d} \tau$ denote both integration and summation over spatial and spin degrees of freedom.  
Multiplying the SE on both sides with a smooth and compactly supported function $\phi\in \mathcal{C}^{\infty}_\mathrm{c}(\Omega^N)$, a so-called {\it test function}, and integrating by parts yields ($\nabla = (\nabla_{r_1}, \dots, \nabla_{r_N})$)
\begin{equation}
\label{eq:WeakSe}
\frac{1}{2}\int_{\Omega^N}\nabla \psi \cdot \nabla\phi~\mathrm d\tau
+
\int_{\Omega^N}\psi \, \hat V_\mathrm{C} \, \phi~\mathrm d\tau
=
E_0\int_{\Omega^N}\psi  \phi~\mathrm d\tau
~,
\end{equation}
where $\hat V_\mathrm{C}$ denotes the Coulomb operator (containing both the Coulomb attraction and repulsion) and $\psi$ a solution of the SE. 
It follows immediately that the l.h.s. of Eq.~\eqref{eq:WeakSe} defines a bilinear form $a(\cdot , \cdot):\mathcal{C}^{\infty}_\mathrm{c}\times \mathcal{C}^{\infty}_\mathrm{c} \to \mathbb{K}$ with $\mathbb{K}$ being the underlying algebraic field. 
Boundedness and ellipticity of this bilinear form, however, are non-trivial consequences that go back to Hardy--Rellich inequalities proving that $\hat V_\mathrm{C}:\mathcal{C}^{\infty}_\mathrm{c}\to L^2$ is bounded (for a general introduction see \cite{yafaev1999sharp}).
Note that this treatment of the SE extends the set of admissible wave functions to the set of antisymmetric $L^2$-functions $\psi$ of finite kinetic energy $\mathcal K(\psi)$, i.e.,
\begin{align*}
\Vert \psi \Vert_2^2  :=\int_{\Omega^N}|\psi|^2\mathrm d\tau <+\infty
\end{align*}
and
\begin{align*} 
\mathcal K(\psi):=\sum_{i=1}^N\int_{\Omega^N}|\nabla_{r_i} \psi|^2\mathrm d\tau<+\infty~.
\end{align*}
We denote this space $H^1(\Omega^N)$ and impose the norm
\begin{align*}
\Vert \cdot \Vert : H^1\to \mathbb{R}~;~ \psi \mapsto \sqrt{\Vert \psi \Vert_2^2 + \mathcal K(\psi)}~.
\end{align*}
In this topology $\mathcal{C}^{\infty}_\mathrm{c}(\Omega^N)\subseteq H^1(\Omega^N)$ is dense.
Hence, the bilinear form $a(\cdot, \cdot)$ is continuously extendable to $H^1(\Omega^N)$.
We define the operator
\begin{equation*}
\hat H_\mathrm{w}:H^1\to (H^1)'~;~\psi \mapsto \hat H_\mathrm{w}\psi = a(\psi, \cdot)~,
\end{equation*}
where $(H^1)'$ is dual space of $H^1$, which we shall denote $H^{-1}$ from now on.
Note that $\hat H_\mathrm{w}$ maps indeed into $H^{-1}$ since boundedness and ellipticity are preserved under continuous extensions.
Furthermore, the r.h.s. of Eq.~\eqref{eq:WeakSe} can be generalized to the {\it dual pairing} allowing to reformulate the SE as operator equation: Find $\psi \in H^1$ such that $\hat H_\mathrm{w} \psi = E_0 \psi'$, with $\psi'$ being the {\it Riesz representation} of $\psi$.
This general approach to the SE was to the best of our knowledge not considered in the mathematical analyses of CC theory prior to the work of Schneider and Rohwedder \cite{Schneider2009analysis,Rohwedder2013continuous,Rohwedder2013error}.
Subsequently, we consider this weak formulation and for simplicity write $\hat H_\mathrm{w} = \hat H$.
\newline
\indent Different parameterizations of $\psi$ lead to different approximation schemes, subject of this article is the CC scheme, i.e., we parameterize $\psi$ on an exponential manifold.
We assume that the solution $\psi_*$ can be written $\psi_* = \phi_0 + \psi_\perp$, where $\phi_0$ is a reference determinant of $N$ one-electron functions and $\psi_\perp$ is an element of $\{\phi_0\}^{\perp}$, the $L^2$-orthogonal complement of $\phi_0$. 
We denote the $L^2$-inner product by $\langle \cdot \vert \cdot \rangle$ and follow the quantum chemistry notation for expectation values of operators, i.e., $\langle \psi \vert \hat A \vert \psi\rangle $. 
In particular, assuming that $\hat H$ supports a ground state, which is always the case for Coulomb systems \cite{yserentant2010regularity}, the Rayleigh--Ritz variational principle reads
\begin{align*}
E_0
= \min_{ \psi\neq 0}\frac{\langle \psi |\hat H |\psi \rangle}{\langle \psi | \psi \rangle}
=: \min_{ \psi \neq 0} \mathcal{R}(\psi)~,
\end{align*}
with $\psi \in H^1$.
Note that although we assume $\psi$ to be normalizable ($L^2$-summable), we do not impose $\Vert \psi\Vert_2 =1$, but rather $\Vert \phi_0\Vert_2 =1$. 
Furthermore, by construction of the solution $\psi_*$, we assume intermediate normalization $\langle\phi_0|\psi \rangle=1$. \newline
\indent Next, let $\{\chi_k\}\subset H^1(\Omega)$ be an $L^2(\Omega)$-orthonormal one-electron basis of the space of admissible one-electron wave functions. 
Unless we explicitly write $\{\chi_k\}_{k=1}^K$ we refer to the infinite dimensional setting.
We construct from this set an $L^2(\Omega^{N})$-orthonormal {\it Slater basis} in the usual fashion denoted $\{\phi_\mu\}$. 
Note that the $N$-particle basis functions $\{\phi_\mu\}$ span the infinite dimensional space of all possible excitations with respect to the reference determinant $\phi_0$.
In this notation we have $\psi = \phi_0 + \psi_\perp= \phi_0 + \sum_{\mu} s_\mu \phi_\mu$, where $\{s_\mu\}$ are the $L^2$-weights of $\psi$ in the given Slater basis, i.e, $s_\mu = \langle\phi_\mu\vert \psi\rangle$. 
We formally define the cluster operators by $\hat S = \sum_{\mu} s_\mu \hat X_\mu$, where $\hat X_\mu$ excites the reference state $\phi_0$ to the state $\phi_\mu$. We obtain $\psi=(\hat I +\hat S)\phi_0$ with $\hat I$ denoting the identity operator. 
The coefficients $s_\mu$ are called {\it cluster amplitudes} and we say that $s=\{s_\mu\}$ is a set of admissible cluster amplitudes if $\hat S\phi_0\in L^2$ and $\mathcal K(\hat S\phi_0)<+\infty$. 
Due to the one-to-one relationship between cluster amplitudes and linearly parametrized wave functions, a natural choice for a norm on the space of admissible cluster amplitudes is the corresponding wave function norm of $\hat S \phi_0$~\cite{Schneider2009analysis,Rohwedder2013continuous}, i.e.,
\begin{align*}
\Vert s \Vert^2 =\Vert \hat S\phi_0 \Vert^2 =  \Vert \hat S\phi_0 \Vert_2^2 + \mathcal K(\hat S\phi_0)~.
\end{align*}

\subsection{The Exponential Ansatz}
\label{Sec:ExpAn}
The CC theory is based on an exponential parametrization of wave functions.
This is an alternative and, assuming full {\it excitation rank} (explained below) of the cluster operators, equivalent description of the full CI (FCI) wave function. 
Since its introduction by Hubbard \cite{hubbard1957description} and, independently, Hugenholtz \cite{hugenholtz1957perturbation}, the unique parametrization of a wave function $\psi$ by the exponential $\psi = e^{\hat T}\phi_0$ was assumed to be true and motivated from formal manipulations.
However, the unique representation of functions in a Hilbert space is by nature a mathematical problem and was rigorously proven for the exponential parametrization in the infinite dimensional case by Rohwedder~\cite{Rohwedder2013continuous}. \newline
\indent A key element in deriving the exponential parameterization from the mathematical viewpoint is the well-definedness of the exponential of $\hat T$ (or equivalently the logarithm of $\hat I+\hat S$), which is subject of functional calculus.
We emphasize that the applicability of functional calculus depends strongly on the operator's domain since different domains may imply different properties of the operator, e.g., boundedness, essential self-adjointness, sectorial spectrum, etc.
By the fact that Rohwedder \cite{Rohwedder2013continuous} showed the $H^1$-continuity of cluster operators in a continuous setting, the functional calculus for bounded operators was proven to be applicable. \newline
\indent In the finite dimensional case this result was known in the quantum chemistry community, see e.g. \v{Z}ivkovi{\'c} and Monkhorst~\cite{vzivkovic1978analytic}.
However, this result was revisited by Schneider \cite{Schneider2009analysis} using  the {\it Cauchy--Dunford} calculus.
To the best of our knowledge, the subtleties addressed in \cite{Schneider2009analysis,Rohwedder2013continuous} have not been part of previous considerations in mathematical analysis of CC theory. 
These important results demonstrate how quantum chemistry benefits from mathematics on a very fundamental level.
The continuous CC theory amounts to the exact formulation where the set $\{\chi_k\}$ forms a basis (in the strict mathematical sense) of the one particle space  $H^1(\Omega)$.
In a for this article appropriate form, we recall Rohwedder's result~\cite{Rohwedder2013continuous}:

\textit{(i) Let $\phi_0$ denote a reference determinant, e.g., the Hartree--Fock solution.
	Given a wave function $\psi_{\perp}\in \{ \phi_0\}^\perp \cap L^2$, i.e., $\langle \psi_{\perp} | \phi_0 \rangle = 0 $, 
	set $S=S_{\psi_{\perp}}$ where $S_{\psi_\perp} \phi_0 = \psi_\perp$ and note that $S\in \mathcal{B}(L^2,L^2)$, i.e., a bounded linear operator from $L^2$ into $L^2$.
	Then, $\psi_{\perp} \in H^1$ if and only if $S\in \mathcal{B}(H^1,H^1)$. 
	Furthermore, there exists a constant $C$ independent of $\psi_\perp$ such that
	\[
	\Vert \psi_{\perp} \Vert \leq \Vert S\Vert  \leq C \Vert \psi_{\perp} \Vert~.
	\]
	An equivalent statement holds for the $L^2$-adjoint of $S$.\newline
	\indent (ii) The exponential map $\hat T\mapsto e^{\hat T}$ is a $C^\infty$ isomorphism between $\mathcal C := \{ \hat T: \hat T \in \mathcal{B}(H^1,H^1)  \}$ and 
	$\mathcal I + \mathcal C := \{\hat I+  \hat T: \hat T \in \mathcal{B}(H^1,H^1)  \}$. In particular, for any $\psi\in H^1$ with $\langle \phi_0\vert \psi\rangle=1$ there exists a unique $\hat T$ such that $\psi= e^{\hat T}\phi_0$.} \newline
\indent Note that this result holds for any orthonormal set of $N$-particle basis functions spanning the space of selected excitations with respect to the reference determinant $\phi_0$. 
However, it is required that the excitation rank of the cluster operators remains untruncated, i.e., $\hat T = \sum_{k=1}^N \hat T_k$, where $\hat T_1$ corresponds to single excitations, $\hat T_2$ to double excitations, ..., $\hat T_N$ to $N$-fold excitations.
Consequently, we have 
\begin{align}
\label{eq:ExpTN}
\psi = \exp(\hat T_1 + \dots + \hat T_N)\phi_0
\end{align}
in the case of full excitation rank. \newline
\indent The usual identification between the linear and exponential parametrization holds~\cite{helgaker2014molecular}:
Write $\hat S = \hat S_1 + \dots +\hat S_N$ and suppose that the linear parametrization is given by 
\begin{align}
\label{eq:CISN}
\psi = (\hat I + \hat S_1 + \dots +\hat S_N)\phi_0~.
\end{align}
Expanding the exponential in Eq.~\eqref{eq:ExpTN}, and comparing with Eq.~\eqref{eq:CISN}, then yields
\begin{align*}
\hat T_1 = \hat S_1,\quad \hat T_2 = \hat S_2  - \frac 1 2 \hat S_1^2,\quad ...
\end{align*}
and for the amplitudes
\begin{align*}
t_i^a = c_i^a/c_0,\quad t_{ij}^{ab} = c_{ij}^{ab}/c_0-(c_i^ac_j^b-c_j^ac_i^b)/c_0^2,\quad ... \quad ,
\end{align*}
where $c_0$ is the FCI coefficient of the reference determinant (here $c_0=1$).
This shows a one-to-one relation for untruncated linear and exponential parameterizations.
Restricting the parametrization on the sub-manifold of excitation rank $k<N$, this one-to-one relationship is in general not true (see Remark~2 in \cite{laestadius2017analysis}): 
Consider CCSD for $N>2$ particles, i.e., $\psi=e^{\hat T_1+\hat T_2}\phi_0$.
Expanding the exponential yields 
\begin{align*}
\hat T_1+\hat T_2+\frac{(\hat T_1+\hat T_2)^2}{2}+...+\frac{(\hat T_1+\hat T_2)^N}{N!}=\hat S~,
\end{align*}
which is not a CISD parametrization, unless for the trivial case $\hat T_1= \hat T_2= 0$.

\subsection{The Coupled-Cluster Energy}
\label{sec:Inc}
Being able to express any wave function in $H^1$ on an exponential manifold, it is straightforward to derive the linked CC equations \cite{helgaker2014molecular}:
\begin{align}\label{eq:CCeqs}
\left\lbrace
\begin{aligned}
\mathcal E(t) &= \langle \phi_0 | e^{-\hat T} \hat H e^{\hat T} \vert \phi_0 \rangle ~,\\
(f(t) )_\mu &= \langle \phi_\mu | e^{-\hat T} \hat H e^{\hat T} \vert \phi_0 \rangle = 0~,~ \text{for all } \phi_\mu~.
\end{aligned}
\right.
\end{align}         
Here, $\phi_0$ and all the (visavi $\phi_0$) excited determinants $\phi_\mu$ are assumed to form a basis of the anti-symmetric part of $H^1$. 
Note that the above equation defines the CC function $f$ and the CC energy function $\mathcal E$. 
Theorem 5.3 from \cite{Rohwedder2013continuous} demonstrates that the CC theory provides a wave function that satisfies $\mathcal R(\psi_*) = E_0 = \mathcal E(t_*)$:

{\it The continuous (and with full excitation rank) CC amplitudes $t_*$ solve $f(t_*)=0$ fulfilling $\mathcal E(t_*) = E_0$ if and only if the corresponding function $\psi_*= e^{\hat T_*} \phi_0$ solves the SE $\hat H \psi_* = E_0 \psi_*$.}

By this fact and with $E_0 = \mathcal E(t_*)$, if $t_*$ solves $f(t_*)=0$ the SE yields
\[
\langle \psi_* | \hat H |\psi_* \rangle = E_0 \langle \psi_*| \psi_*\rangle~.
\] 
Hence, the CC amplitudes describe a function $\psi_*$ that provides the system's energy in the usual quantum mechanical setting, i.e., $\mathcal R(\psi_*) =  \mathcal E(t_*)$. \newline
\indent In practice, computations are carried out using a finite basis $ \{ \chi_k \}_{k=1}^K$ and furthermore with a truncated excitation rank $\hat T^{(n)} =\hat T_1 + \dots \hat T_n$, $n<N$. 
The total truncation level can then be denoted by $d=(K,n)$, and where we solve $f=0$ on $\mathcal V^{(d)}$ to obtain $f(t_d)=0$, $E_d = \mathcal E (t_d)$.
We note the following from the literature: \newline
\indent (i) Given a finite one-electron basis $\{ \chi_k\}_{k=1}^K$, we denote the span of the corresponding Slater basis by $H_K^1$. With full excitation rank ($n=N$) Proposition~4.7 in~\cite{Schneider2009analysis} gives: $f(t_d)=0$ and $E_K = \mathcal E(t_d)$ if $\psi_d =e^{\hat T_d} \phi_0$ solves the SE on $H_K^1$, i.e., $\hat H e^{\hat T_d} \phi_0= E_K e^{\hat T_d} \phi_0$. 
By the argument of Monkhorst in~\cite{monkhorst1977calculation} we can establish the reverse: Assume $f(t_d)=0$ and set $E_K = \mathcal E(t_d)$, then since $\hat I_K = \vert \phi_0\rangle \langle \phi_0 \vert + \sum_\mu \vert\phi_\mu\rangle \langle \phi_\mu \vert$ we obtain (Eq.~(38) in~\cite{monkhorst1977calculation})
\begin{align*}
&\langle \phi_0\vert e^{\hat T_d^\dagger} e^{\hat T_d}\vert \phi_0\rangle \mathcal R(e^{\hat T_d} \phi_0) 
=  \langle \phi_0\vert e^{\hat T_d^\dagger} e^{\hat T_d}\hat I_Ke^{-\hat T_d} \hat H e^{\hat T_d}\vert \phi_0\rangle \\
&= \langle \phi_0\vert e^{\hat T_d^\dagger} e^{\hat T_d}\vert \phi_0\rangle  \mathcal E(t_d)  
+ \sum_\mu \langle \phi_0\vert e^{\hat T_d^\dagger} e^{\hat T_d}\vert \phi_\mu\rangle f(t_d)~.
\end{align*}
From this we can conclude $\mathcal R(e^{\hat T_d} \phi_0) = \mathcal E(t_d) = E_K$, i.e., the CC wave function gives the energy when inserted into the Rayleigh--Ritz quotient. Furthermore, we have (where $\mathcal C_d$ denotes the truncated version of $\mathcal C$)
\begin{align*}
&\inf\{ \mathcal R(\psi) : \psi = e^{\hat T}\phi_0, \hat T \in \mathcal C_d \}\\
&= \inf\{ \mathcal R(\psi) : \psi = (\hat I + \hat S)\phi_0, \hat S \in \mathcal C_d \} = E_K~,
\end{align*}
by the equivalence between linear and exponential parametrization as long as full excitation rank is kept. Consequently, $\psi_d = e^{\hat T_d} \phi_0$ solves the SE on $H_K^1$, which establishes the reversed implication in Proposition~4.7 in~\cite{Schneider2009analysis}. \newline
\indent (ii) However, for $n<N$ we have in general (see for instance Remark~4.9 in~\cite{Schneider2009analysis})
\begin{align*}
E_d^\mathrm{var}:= \inf\{ \mathcal R(\psi) : \psi = e^{\hat T}\phi_0, \hat T \in \mathcal C_d \} \neq E_d, \quad n<N~,
\end{align*}
which gives the well-known result that the computed $E_d$ is not an upper bound to $E_K$. Hence, $E_d \neq \mathcal R(e^{\hat T_d} \phi_0)$ where $f(t_d)=0$ and $E_d = \mathcal E(t_d)$. \newline
\indent (iii) By (ii), strictly speaking, CC methods do not compute wave functions, as $\psi_d$ does not provide the system's energy and therewith does not fulfill the Copenhagen interpretation's first principle~\cite{heisenberg1963kopenhager}.
However, as mathematical analyses in \cite{Schneider2009analysis,Rohwedder2013continuous,Rohwedder2013error,laestadius2017analysis,faulstich2018analysis} have demonstrated, CC methods do provide approximate wave functions that converge to the solution of the SE (as $K\to\infty$, $n\to N$).  
The Copenhagen interpretation is formulated for full systems, which correspond to the continuous CC formulation, and does not contain any statement about approximative solutions. 
This raises the fundamental question of what properties should be demanded for approximative solutions. \newline
\indent (iv) To contrast with the next section, we would also like to point out the work \cite{vzivkovic1978analytic} where, for a finite basis, the CC equations were analyzed in a perturbational setting. Writing 
\begin{align*}
e^{\hat T^{(n)}} & =  \hat I + \hat T_1 + \dots + \hat T_n + \lambda \big(\hat T_{n+1}(n) + \hat T_{n+2}(n)\big)~, 
\end{align*}
where we followed the notation in \cite{vzivkovic1978analytic} (see Eqs. (A9) and (A10)), the CI equations are obtained at $\lambda=0$ and $\lambda=1$ corresponds to the CC case. 
From this and under the assumption of a finite one-electron basis, both the reality and multiplicity of the CC solutions were investigated with respect to pole and branch cut singularities in the complex plane. 
The emergence of multiple solutions is certainly interesting and worth pursuing, however, 
the local analysis studied here instead deals with the establishment of a \emph{locally} unique solution under certain assumptions.
Note that the local behavior of a solution is important for the applicability and convergence of Newton--Rhapson and quasi-Newton methods.

\section{Local Analysis in Coupled-Cluster Theory}
\label{sec:loc}
The CC equations (linked and unlinked) can be formulated as a \textit{non-linear Galerkin scheme}, which is a well-established framework in numerical analysis to convert the continuous Schr\"odinger equation to a discrete problem.
Instead of solving the full problem, Galerkin methods solve the CC equations in a finite dimensional subspace $H_{d}\subseteq H^1$.
Note that the CC equations remain the same, only the space spanned by the considered $\{ \phi_\mu\}$ has changed. 
Reducing the problem to a finite-dimensional vector subspace allows to numerically compute an approximate solution via Newton--Rhapson or quasi-Newton methods. 
Galerkin methods allow a local analysis, which is useful for CC theory due to the manifold of solutions \cite{vzivkovic1977existence,vzivkovic1978analytic,jeziorski1989valence,piecuch1990coupled,kowalski1998towards,piecuch2000computational} and the use of quasi-Newton methods that require certain local behavior of the solutions.
Local analysis furthermore allows reliable statements about the existence and local uniqueness of Galerkin solutions as well as quantitative statements on the basis-truncation error. 
Its backbone is formed by a local version of Zarantonello's theorem~\cite{zaidler1990nonlinear}:\newline
\indent
{\it 
	Let $f:X \to X'$ be a map between a Hilbert space $(X,\langle\cdot,\cdot\rangle,\Vert \cdot \Vert)$ and its dual $X'$, and let $x_*\in B_{\delta}$ be a root, $f(x_*)=0$, where $B_{\delta}$ is an open ball of radius $\delta$ around $x_*$.
	Assume that f is Lipschitz continuous and strongly monotone in $B_{\delta}$ with constants $L > 0$ and $\gamma>0$, respectively. Then  
	the root $x_*$ is unique in $B_{\delta}$. 
	Indeed, there is a ball $C_{\varepsilon}\subset X'$ with $0\in C_{\varepsilon}$ such that the solution map $f^{-1}:C_{\varepsilon}\to X$ exists and is Lipschitz continuous, implying that the equation 
	$f(x_*+x)=y$
	has a unique solution $x=f^{-1}(y)-x_*$, depending continuously on $y$, with norm $\Vert x \Vert \leq \delta$.
	Moreover, let $X^{(d)}\subset X$ be a closed subspace such that $x_*$ can be approximated sufficiently well, i.e., the distance $d(x_*,X^{(d)})$ is small. 
	Then, the projected problem $f_d(x_d)=0$ has a unique solution $x_d\in X^{(d)} \cap B_{\delta}$ and 
	\begin{equation*}
	\Vert x_*-x_d \Vert\leq \frac{L}{\gamma} d(x_*,X^{(d)})~,
	\end{equation*}
	i.e., $x_d$ is a quasi-optimal solution.}\newline
\indent The concept of quasi optimality was introduced in Jean C\'ea's dissertation \cite{cea1964approximation} in 1964 for linear Galerkin schemes and got extended over the years to the non-linear case.
It ensures that the Galerkin solution in a fixed approximative space is, up to a multiplicative constant, the closest element to the exact solution.      
For obvious reasons this is a desired property for CC schemes. 
The different CC methods vary, however, in more than just minor details, which makes this property a conceptual different and challenging task to establish for each method. 

\subsection{Local Unique Solutions and Quasi-Optimality}
\label{sec:locDefs}
We start by elaborating on the assumptions of Zarantonello's theorem in a more demonstrative way. 
Here, the notation $\langle s , t\rangle = \sum_{\mu}s_\mu t_\mu$ is used for sequences $s = \{s_\mu\}$ and $t = \{t_\mu\}$.
In the context of the CC theory, the CC function $f$ from Eq.~\eqref{eq:CCeqs} is said to be {\it strongly monotone} if for sets of cluster amplitudes $t = \{t_\mu\}$ and $t'= \{t_\mu'\}$ there exists a $\gamma>0$ such that 
\begin{align}\label{eq:StrongM}
\langle f(t)-f(t'), t-t'\rangle \geq\gamma \Vert t-t'\Vert^2~.
\end{align} 
If this inequality is true for all $t,t'\in B_\delta(t_*)$ then $f$ is said to be locally strongly monotone.
The CC function $f$ is further said to be {\it Lipschitz continuous} if there exists a constant $L>0$ such that
\begin{align}
\label{eq:LipDef}
\Vert f(t) - f(t')\Vert \leq L \Vert t-t'\Vert~.
\end{align}
In direct analogy with local strong monotonicity, we define local Lipschitz continuity if Eq.~\eqref{eq:LipDef} is fulfilled for all cluster amplitudes $t,t'$ inside some ball.\newline
\indent To exemplify these concepts in a simple way we consider a smooth function ${f:\mathbb{R}\to\mathbb{R}}$.
By the Cauchy--Schwarz inequality, the strong monotonicity implies that the derivative $f'(t)\geq \gamma$, i.e., $f$ is a strictly monotonically increasing function.
Note that strictly monotone functions are injective (one-to-one), which implies local invertibility.
Hence, this already ensures local uniqueness of the function's root $t_*$, if supported.
Lipschitz continuity on the other hand implies that $-L\leq f'(t)\leq L$.
Hence, the assumptions in Zarantonello's theorem are restrictions to the function's slope, namely 
\[
0< \gamma\leq f'(t)\leq L~.
\]
By introducing normed operator spaces, these restrictions can be generalized to vector valued and even infinite dimensional functions $f$. \newline
\indent Returning to the general case, the Lipschitz continuity is key to derive the quasi-optimality in case of Galerkin solutions. 
We assume that $X^{(d)}\subsetneq X$ is the considered approximation space supporting the Galerkin solution $t_d$, i.e., $\langle f(t_d),s\rangle=0$ for all $s\in X^{(d)}$. 
Then, $f(t_*)-f(t_d)\in(X^{(d)})^{\perp}$, i.e., $\langle f(t_*)-f(t_d) , u \rangle = 0$ for all $u\in X^{(d)}$, in particular for $u=t_d$.
Starting from the strong monotonicity, we deduce for any $u\in X^{(d)}$ that
\begin{align*}
\gamma \Vert t_*-t_d\Vert ^2
&\leq 
\langle f(t_*)-f(t_d),  t_*-t_d\rangle\\
&=
\langle f(t_*)-f(t_d),  t_*-u\rangle\\
&\leq 
L\Vert t_*-t_d\Vert \Vert  t_*-u\Vert~.
\end{align*}  
Because $u\in X^{(d)}$ was chosen arbitrarily, the above estimate holds for all $u$, which implies the quasi optimality:
\begin{align}
\label{eq:QO}
\Vert t_*-t_d\Vert  \leq L/\gamma \min_{u\in X^{(d)}} \Vert t_*-u \Vert~.
\end{align}

To apply Zarantonello's theorem to CC methods, the main challenge is to demonstrate a strictly positive $\gamma$ in Eq.~\eqref{eq:StrongM} such that strong monotonicity holds locally around the solution that corresponds to the ground state. 
The original idea in \cite{Schneider2009analysis} to obtain such a result in the finite-dimensional projected CC theory assumed the existence of a HOMO-LUMO gap. Further, more technical assumptions on the Fock operator $\hat F$ (see G\aa rding inequality below) were needed to achieve a generalization to the continuous CC setting~\cite{Rohwedder2013error}, which also has a counterpart for $\hat H$. We refer the reader to \cite{Schneider2009analysis,Rohwedder2013continuous,Rohwedder2013error,laestadius2017analysis,faulstich2018analysis} for the detailed proofs and made assumptions, not only within the traditional CC formalism,  but also for the TCC and extended CC methods.
However, we remark that these assumptions are sufficient conditions but not necessary.
One example is given by metals: Despite their typically small or negligible HOMO-LUMO gaps, the single-reference CC method can compute metallic effects often quite well.
This suggests that the HOMO-LOMO gap assumption, which limits the results' applicability, can be lifted in the case of non-multi-configuration systems \cite{faulstich2018analysis}.
See also~\cite{jeziorski1989valence} for a CC theory that considers open-shell systems where no HOMO-LUMO gap exists.
\newline
\indent Here, we extend the results in \cite{Schneider2009analysis,Rohwedder2013continuous,Rohwedder2013error,laestadius2017analysis,faulstich2018analysis} by optimizing the strong monotonicity constant $\gamma$, which yields lesser restrictions on the solution's cluster amplitudes $t_* = \{ (t_*)_\mu \}$. Further investigations need to be undertaken before the presented analysis can lead to practical results of the reliability of the CC approach. However, we suggest an estimate on the CC amplitudes that is sufficient to guarantee the existence of a locally unique CC solution (see Eq.~\eqref{eq:Conj1}) and contrast it with the single amplitudes diagnostic of~\cite{Lee1989}.

\subsection{Local Strong Monotonicity of The Coupled-Cluster Function}
\label{sec:strongmon}
In the literature there are two different proofs that the infinite dimensional (continuous) CC function $f$ is locally strongly monotone~\cite{Rohwedder2013error}  (see also~\cite{laestadius2017analysis} for the extended CC function).
Even though spectral-gap assumptions enter the arguments, it is the so-called G\aa rding constants that give a sufficient condition for the local strong monotonicity, as will be demonstrated  below.  
This fact emerges from the analysis in \cite{Rohwedder2013error} but was noted and elaborated within the analysis of the extended CC method in \cite{laestadius2017analysis}. We here furthermore improve the existing analysis by optimizing the constants.
We start by defining the G\aa rding inequality that will be used extensively in the sequel: 

{\it An operator $\hat A$ fulfills a G\aa rding inequality if there exists a real constant $e$ such that $\hat A+e$ is coercive, i.e., there exists a constant $c>0$ that depends on $e$ (we denote this dependence by $c(e)$) such that
	\[
	\langle \psi\vert \hat A+e \vert\psi\rangle \geq c(e)\Vert \psi\Vert^2~.
	\]}

The coercivity above describes a particular growth behavior of $\hat A+e$ as the lower bound becomes large when the wave function is at the extreme of the space, e.g., wave functions with a large kinetic energy.
Subsequently, we denote the l.h.s. of Eq.~\eqref{eq:StrongM} by $\Delta$, i.e., for two sets of CC amplitudes $t = \{t_\mu\}$ and $t'= \{t_\mu'\}$ we have
\[
\Delta = \langle f(t)-f(t'), t-t'\rangle~.
\] 
We further set $\Delta \hat T = \hat T-\hat T'$, which yields by the CC equations in Eq.~\eqref{eq:CCeqs} the equality
\begin{align}\label{eq:Delta}
\Delta = \langle \Delta \hat T \phi_0| e^{-\hat T}\hat H e^{\hat T}-e^{-\hat T'}\hat H e^{\hat T'}|\phi_0\rangle~.
\end{align}

Next, we elaborate on G\aa rding inequalities for two different operators that imply local strong monotonicity of the CC function, by bounding the r.h.s. of Eq.~\eqref{eq:Delta}. 
Interestingly, for the finite-dimensional (projected) CC method, only the latter approach has a counterpart (using the particular structure of the Fock operator $\hat F$).

\subsubsection{A G\aa rding Inequality for the Hamiltonian}
\label{sec:GardingH}
We here assume a spectral gap $\gamma_*$ of $\hat H$, i.e., for all $\psi$ that are $L^2$-orthogonal to the ground state $\psi_*$ we have
$\mathcal R(\psi) - E_0 \geq \gamma_*$, for some $\gamma_*>0$, i.e., we assume a non-degenerate ground state.
We also assume that $\phi_0$ is a good approximation of the exact wave function, i.e., $\varepsilon = \Vert\psi_*-\phi_0 \Vert_2$ is small. 
It then holds (see Lemma~11 in \cite{laestadius2017analysis})
\begin{equation}
\label{eq:HspecGAP}
\langle \hat T\phi_0| \hat H-E_0| \hat T\phi_0\rangle \geq \gamma_*(\varepsilon) \Vert \hat T \phi_0 \Vert_{2}^2~,
\end{equation}
with $\gamma_*(\varepsilon) = \gamma_*(1 -4\varepsilon + \mathcal O(\varepsilon^2))$.
Thus, $\gamma_*(\varepsilon)$ is close to $\gamma_*$ and  strictly positive, if $\varepsilon$ is sufficiently close to zero.  
Using the argument in~\cite{Rohwedder2013error,laestadius2017analysis} (see proof of Theorem~3.4 in~\cite{Rohwedder2013error}, and also Eq.~(16) with $\hat \Lambda_*=0$ together with the proof of Theorem~16 in~\cite{laestadius2017analysis}), we obtain
\begin{equation}
\begin{aligned}
\label{eq:Delta1}
\Delta &\geq \langle \Delta\hat  T\phi_0|\hat H-E_0|\Delta \hat  T\phi_0\rangle \\
&\quad -\left( 
\Vert e^{-\hat T_*^{\dagger}}-\hat I\Vert
+\Vert e^{-\hat T_*^{\dagger}}\Vert \Vert e^{\hat T_*}-\hat I\Vert\right)\Vert \Delta \hat T \phi_0\Vert^2~.
\end{aligned}
\end{equation}
In \cite{Rohwedder2013error}, the first term of Eq.~\eqref{eq:Delta1} was bounded by a constant times $\Vert \Delta \hat T \phi_0\Vert^2$, achieved by combining the G\aa rding inequality with Eq.~\eqref{eq:HspecGAP}. \newline
\indent From Lemma~11 in \cite{laestadius2017analysis}, it follows that
\begin{align*}
\langle \Delta\hat T\phi_0|\hat H-E_0|\Delta\hat T\phi_0\rangle 
\geq 
\frac{\gamma_*(\varepsilon)}{\gamma_*(\varepsilon) + e+E_0} c(e)\Vert \Delta\hat T\phi_0 \Vert^2~.
\end{align*}
However, this can be further strengthened to   
\begin{align*}
\langle \Delta\hat T\phi_0|\hat H-E_0|\Delta\hat T\phi_0\rangle 
\geq 
\eta_\mathrm{opt}(\varepsilon)\Vert \Delta\hat T\phi_0 \Vert^2~,
\end{align*}
with the optimal constant
\[
\eta_\mathrm{opt}(\varepsilon):= \max_{e>0}\frac{\gamma_*(\varepsilon)}{\gamma_*(\varepsilon) + e+E_0} c(e)~.
\]
From this we conclude
\begin{align}
\Delta &\geq \Big( \eta_\mathrm{opt}(\varepsilon)
-\Vert e^{-\hat T_*^{\dagger}}-\hat I\Vert\nonumber \\
&\quad -\Vert e^{-\hat T_*^{\dagger}}\Vert \Vert e^{\hat T_*}-\hat I\Vert\Big)\Vert t-t'\Vert^2~,
\label{eq:DeltaEta}
\end{align} 
which yields the following {\it sufficient condition} for the local strong monotonicity of $f$, namely
\begin{align}
\label{eq:crit1}
\eta_\mathrm{opt}(\varepsilon) > \Vert e^{-\hat T_*^{\dagger}}-\hat I\Vert
+\Vert e^{-\hat T_*^{\dagger}}\Vert \Vert e^{\hat T_*}-\hat I\Vert~.
\end{align}
Given $\gamma_*>0$, we observe that a sufficiently small $\varepsilon$ and $t_*$, such that $\Vert \hat T_* \Vert$ is small enough relative to  $\eta_\mathrm{opt}(\varepsilon)$, guarantees that Eq.~\eqref{eq:crit1} is fulfilled. (Recall that $\Vert t \Vert $ and $\Vert \hat T \Vert$ are equivalent, see Section~\ref{Sec:ExpAn}.)  \newline
\indent To finalize this section, we offer the following interpretation of Eq.~\eqref{eq:DeltaEta}, providing a more descriptive approach to Eq.~\eqref{eq:crit1}. We see as $e$ tends to $-E_0$ from above, the quotient ${\gamma_*(\varepsilon)}/({\gamma_*(\varepsilon) + e+E_0})$ goes to one from below.
Furthermore, assume that $c(e)$ goes to zero from above as  $e$ approaches $ -E_0$ from above.
This suggest an optimal value of $e_\mathrm{opt}>-E_0$.
For instance, choosing $e_n=-E_0+\gamma_*(\epsilon)/n$ implies 
\[
\frac{\gamma_*(\varepsilon)}{\gamma_*(\epsilon) + e+E_0}c(e_n)
=
\frac{1}{1+1/n} c(e_n)
\]
such that $\gamma_*(\varepsilon)$ is eliminated from the expression. 
Assuming further that $e_\mathrm{opt}$ corresponds to an $n_\mathrm{opt}\gg 1$ yields $\eta_\textrm{opt}\approx c(e_\mathrm{opt})$. In conclusion, as long as $\gamma_*(\varepsilon)>0$, the G\aa rding constant $c(e_\mathrm{opt})$ offers a direct estimate of the monotonicity constant $\gamma\approx c(e_\mathrm{opt}) -  2 \Vert\hat  T_*\Vert + \mathcal O(\Vert \hat T_*\Vert^2)$. We therefore obtain the following (approximate) sufficient condition for local strong monotonicity 
\begin{equation}
\label{eq:Conj1}
c(e_\mathrm{opt}) >  2\Vert T_*\Vert~.
\end{equation}
Note that $\Vert \hat T \Vert \geq  K \Vert t \Vert$, for some constant $K$. However, a sharp estimate for this constant is object of current research.
Thus, for Zarantonello's theorem to guarantee a locally unique solution, the exact amplitudes $t_*= \{(t_*)_\mu\}$ cannot be too large relative to $c(e_\mathrm{opt}) $.
We remark that by an appropriate choice of the reference determinant $\phi_0$, the single amplitudes $t_1=\{ (t_1)_\mu\}$ do not contribute to (the overall) $\Vert t \Vert$. Thus, if $\Vert t \Vert$ is too large then this is a consequence of $t_2,t_3, \dots$ (doubles, triples, etc.). Numerical investigations are left for future work but 
we can already compare this mathematically derived sufficient condition for locally unique CC solutions with the $t_1$-diagnostics of~\cite{Lee1989}. Given the truncation level $n$ of the excitation rank, here the proposed diagnostic uses all cluster amplitudes $t_1,t_2,\dots,t_n$ and not just the single amplitudes $t_1$. This is a clear advantage since, as mentioned above, orbital rotations can be used to rotate out the single amplitudes. However, our diagnostic offers only a sufficient and not a necessary criterion for a local unique solution, i.e., for large $t_2,t_3,\dots$ the current diagnostic is agnostic about local uniqueness and only states that local strong monotonicity cannot be inferred from this particular analysis. We hope that future work will clarify the situation further.

\subsubsection{A G\aa rding Inequality for the Fock Operator}
\label{sec:GardingF}
On the other hand, assume a HOMO-LUMO gap ${\gamma_0>0}$ of the Fock operator $\hat F$ and that  $\phi_0$ is the Hartree--Fock solution, i.e., $\hat F\phi_0=\Lambda_0 \phi_0$ with 
\[
\langle \psi \vert \hat F-\Lambda_0 \vert \psi\rangle \geq \gamma_0 \Vert \psi \Vert_2^2, \quad \text{for all $\psi \perp \phi_0$}~.
\]
The HOMO-LUMO gap thus corresponds to a spectral gap of the Fock operator and we regard $\Lambda_0$ as the ground-state energy of $\hat F$. 
Let $\hat F = \sum_{i=1}^N \hat f(r_i) $ and choose $\{\chi_k\}$ as eigenbasis of $\hat f$, i.e., $\hat f\chi_k = \lambda_k \chi_k$ for all $k$. We observe that $\Lambda_0 = \sum_{i=1}^N\lambda_i$, $\gamma_0=\lambda_{N+1}-\lambda_N>0$ and $\hat F\phi_\mu = (\Lambda_0+\varepsilon_\mu)\phi_\mu$ with $\varepsilon_{\mu} = \sum_{l\leq |\mu|} \lambda_{a_l}-\lambda_{i_l}$.
The argument proving that the CC function $f$ is locally strongly monotone can then be outlined as follows. \newline
\indent The considered Fock operator is assumed to fulfill a G\aa rding inequality. 
Thus there exists a constant $e$ such that $\hat F+e$ is coercive, i.e.,
\[
\langle \psi|\hat F+e|\psi\rangle \geq c(e)\Vert \psi\Vert^2~.
\]
For the sake of simplicity we use the same symbols for the  G\aa rding constants of $\hat F$ as for the Hamiltonian.
In complete analogy with $\hat H$, the argument in \cite{Schneider2009analysis,laestadius2017analysis} shows that
\begin{align}
\label{eq:FockEst}
\langle \psi|\hat F-\Lambda_0|\psi\rangle 
\geq 
\max_{e>0}\frac{\gamma_0}{\gamma_0 + e+\Lambda_0} c(e)\Vert \psi \Vert^2
\end{align}
and we moreover define
\begin{align}
\label{eq:eatOpt}
\eta_\mathrm{opt}^{(0)} := \max_{e>0}\frac{\gamma_0}{\gamma_0 + e+\Lambda_0} c(e)~.
\end{align}
Following \cite{Rohwedder2013error}, for a fixed $\phi_0$ we define the map from the space of cluster amplitudes into the space of wave functions $O_{\phi_0}:t\mapsto \hat O (t)\phi_0$, with $\hat O:t\mapsto [[\hat F,\hat T],\hat T] + e^{-\hat T} \hat W  e^{\hat T}$.
Hence,
\begin{equation}
\begin{aligned}
\label{eq:HsimFW}
e^{-\hat T}\hat He^{\hat T}\phi_0
&=
e^{-\hat T}(\hat F+\hat W)e^{\hat T}\phi_0\\
&=
(\hat F+[\hat F,\hat T])\phi_0 + \hat O(t)\phi_0~,
\end{aligned}    
\end{equation}
where $\hat H=\hat F+\hat W$, and assume that for some $L>0$ (not too large)
\begin{align} 
\label{eq:LipCond}
\langle \Delta \hat T  \phi_0\vert \hat O(t) - \hat O(t')\vert \phi_0  \rangle  \geq -L \Vert t-t'\Vert^2~.
\end{align}
As a technical remark, the assumption in \cite{Rohwedder2013error} is the stronger requirement that $t \mapsto \hat O(t) \phi_0$ is Lipschitz continuous as a map from the space of cluster amplitudes to $H^{-1}$. 
However, we here note that Eq.~\eqref{eq:LipCond} is sufficient to derive the CC function's local strong monotonicity, as will be evident shortly.
Inserting the identity (a consequence of Eq.~\eqref{eq:HsimFW} and $\hat F \phi_0 = \Lambda_0 \phi_0$)
\begin{align*}
e^{-\hat T}\hat He^{\hat T}\phi_0 = (\hat F + (\hat F-\Lambda_0)\hat T)\phi_0 + \hat O(t)\phi_0
\end{align*}
into Eq.~\eqref{eq:Delta}, as well as using Eqs.~\eqref{eq:FockEst} and \eqref{eq:LipCond}, we obtain
\begin{equation}
\label{eq:strongMont1}
\begin{aligned}
\Delta &= \langle \Delta \hat T\phi_0 \vert \hat F-\Lambda_0\vert \Delta \hat T\phi_0 \rangle\\
&\quad + \langle \Delta\hat  T  \phi_0\vert \hat O(t) - \hat O(t')\vert \phi_0 \rangle\\
& \geq(\eta_\mathrm{opt}^{(0)}  -L) \Vert t - t'\Vert^2~.
\end{aligned}
\end{equation}
Consequently, local strong monotonicity holds if $\eta_\mathrm{opt}^{(0)}>L$. 
Repeating the argument presented in the previous section, with the obvious adaptations, we obtain  
\begin{align}
\label{eq:NonRigFock}
c(e_\mathrm{opt}) > L
\end{align}
as a sufficient condition for $f$ to be locally strongly monotone. Here, no explicit assumption on $\Vert t_* \Vert$ enters. The main drawback of the assumption in Eq.~\eqref{eq:NonRigFock} is that the constant $L$ of the inequality in Eq.~ \eqref{eq:LipCond} has to be determined. Further analysis of this constant is postponed for later work. \newline
\indent Before we conclude this section we exemplify how the G\aa rding constant $c$ can be chosen in the finite dimensional setting. 
In this case the commutator $[\hat F,\hat T]$ is an excitation operator (which implies $[[\hat F,\hat T],\hat T]=0$) and $\hat O(t)$ is simply the similarity transformation of the fluctuation potential $\hat W$. 
This offers the following insight into the optimal constant $\eta_\mathrm{opt}^{(0)}$ in Eq.~\eqref{eq:FockEst} for the truncated case. As in \cite{Schneider2009analysis}, we define the norm on $\{\phi_0\}^{\perp}$ by  
\[
\Vert \hat T\phi_0\Vert_F^2 = \sum_{\mu} \varepsilon_\mu t_\mu^2=\Vert t \Vert_F^2~. 
\]
It follows that
\begin{align*}
\langle \Delta \hat T\phi_0 \vert\hat  F-\Lambda_0\vert \Delta \hat T\phi_0 \rangle = \sum_{\mu} \varepsilon_\mu (\Delta t)_\mu^2 = \Vert \Delta \hat T \phi_0 \Vert_F^2 ~.
\end{align*}
Using $\Vert\hat T\phi_0 \Vert_F$ instead of $\Vert \hat T\phi_0 \Vert$ and making the assumption in Eq.~\eqref{eq:LipCond} also for the truncated theory (denoting the Lipschitz constant in this new topology by $L'$), we obtain
\begin{equation}
\label{eq:EstimationFockNorm}
\begin{aligned}
\Delta &= \sum_{\mu} \varepsilon_\mu (\Delta t)_\mu^2 + \langle  \Delta\hat  T \phi_0\vert  \hat O(t)-\hat O(t')\vert \phi_0 \rangle\\
&= \Vert t-t' \Vert_F^2+\langle  \Delta\hat  T \phi_0\vert  \hat O(t)-\hat O(t')\vert \phi_0 \rangle\\
& \geq (1 - L') \Vert t-t' \Vert_F^2 ~.
\end{aligned}
\end{equation}
Comparing the local strong monotonicity estimates Eq.~\eqref{eq:strongMont1} and Eq.~\eqref{eq:EstimationFockNorm} suggests that the finite-dimensional version of $ \eta_\mathrm{opt}^{(0)}$ equals one.
Furthermore, at first glance it appears that the estimate in Eq.~\eqref{eq:EstimationFockNorm} is obtained without imposing a G\aa rding inequality.
A key observation here is that the choice of the norm makes $\hat F$ on $\{\phi_0\}^{\perp}$ fulfill a G\aa rding inequality with $e_{\mathrm{opt}}=-\Lambda_0$ and $c(e_{\mathrm{opt}})=1$.
Indeed, the inequality is saturated, meaning that equality holds.
It follows then immediately from Eq.~\eqref{eq:eatOpt} that $\eta_{\mathrm{opt}}^{(0)}=c(e_{\mathrm{opt}})=1$.
Thus, in agreement with Eq.~\eqref{eq:NonRigFock} we have obtained the condition $\eta_{\mathrm{opt}}^{(0)}-L'>0$.\newline
\indent To conclude this section, we note that we have formulated an alternative to the diagnostic in Eq.~\eqref{eq:Conj1}: Assume a finite one-electron basis and suppose that $\hat O(t)$ satisfies Eq.~\eqref{eq:LipCond} with the norm $\Vert \cdot\Vert_F$ and $L'<1$ locally around the solution amplitudes. Then local strong monotonicity implies a locally unique CC solution. Whether Eq.~\eqref{eq:LipCond} with $L'<1$ holds without the assumption of a small $\Vert t_* \Vert$ is an interesting and still open question.
Furthermore, the above analysis can be generalized to any single particle operator fulfilling certain properties (see \cite{Schneider2009analysis,faulstich2018analysis}).

\subsection{The Coupled-Cluster Method's Numerical Analysis}
\label{sec:Numer}
As computational schemes, the convergence behavior of CC methods is one of the main objects of study. 
This covers whether or not the method converges towards the exact solution as well as how fast it converges. 
We note that the quasi optimality as given in Eq.~\eqref{eq:QO} yields $t_d\to t_*$ as $d\to \infty$ (for increasing approximation spaces $X^{(d)}$). Furthermore, in the case of the CC method one studies the CC-energy residual 
\[
\vert \mathcal E(t_*)- \mathcal E(t) \vert~.
\]
A major difference between the CI and CC method is that the CC formalism is not variational in the Rayleigh--Ritz sense.
Consequently, it is not evident that the CC energy error decays quadratically with respect to the error of the wave function or cluster amplitudes.
In the sequel we present two approaches that were used in previous mathematical analyses of different CC methods to derive such quadratic error bounds~\cite{Rohwedder2013error,laestadius2017analysis,faulstich2018analysis}. 

\subsubsection{The Aubin--Nitsche Duality Method}
\label{sec:AN}
The Aubin--Nitsche duality method is a standard tool for deriving \textit{a priori} error estimates for finite element methods.
It was introduced independently by Aubin \cite{aubin1967behavior}, Nitsche \cite{nitsche1968kriterium} and Oganesjan--Ruchovets \cite{ruchovec1969study}.
We here elaborate the Aubin--Nitsche duality type method used in \cite{Schneider2009analysis,Rohwedder2013error,faulstich2018analysis} to derive a quadratic error bound for the CC method and the closely related TNS-TCC method, a special case of the tailored CC method \cite{kinoshita2005coupled}.
This approach exploits the mathematical framework introduced by Bangerth and Rannacher \cite{bangerth2013adaptive}. 
The untruncated Euler--Lagrange method gives the Lagrangian $\mathcal{L}(t,s)=\mathcal{E}(t)-\langle f(t),s \rangle$ with $f$ and $\mathcal{E}$ from Eq.~\eqref{eq:CCeqs}.
The corresponding G\^{a}teaux derivative in direction $(u,v)$ is denoted $\mathcal L' (\cdot , \cdot)(u,v)$ and we study $(t_*,s_*)$ fulfilling
\begin{equation}
\label{eq:LagrangeDeriv}
\mathcal{L}'(t_*,s_*)(u,v)
=\left\lbrace
\begin{aligned}
\mathcal{E}'(&t_*)u - \langle f'(t_*)u,s_* \rangle\\
&- \langle f(t_*),v \rangle
\end{aligned}
\right\rbrace=0 ~,
\end{equation}
for all pairs of CC amplitude vectors $(u,v)$.
Under the assumption that $f$ is locally strongly monotone inside a ball around $t_*$, there exists a unique $s_*$ determined by $t_*$ such that $(t_*,s_*)$ solves Eq.~\eqref{eq:LagrangeDeriv}. 
Note, that the assumptions imposed to ensure local strong monotonicity are different for the single-reference CC method \cite{Schneider2009analysis,Rohwedder2013error} and the TNS-TCC method  \cite{faulstich2018analysis}. 
Moreover, there exists a solution $s_d$ to the corresponding discretization of the problem that approximates $s_*$ quasi optimally \cite{Rohwedder2013error,faulstich2018analysis}.
Equipped with these so called {\it dual solutions}, the energy-error characterization given by Bangerth and Rannacher \cite{bangerth2013adaptive} yields
\begin{align*}
2(\mathcal E(t_*)-\mathcal E(t_d))
&=
\mathcal{R}_d^{(3)}+\rho(t_d)(s_*-\upsilon_d)\\
&\quad+\rho^*(t_d,s_d)(t_*-w_d)~,
\end{align*}
with arbitrarily chosen discrete CC amplitudes $\upsilon_d,w_d$.
The given remainder term $\mathcal{R}_d^{(3)}$ is cubic in the {\it primal} and {\it dual} error, i.e., $e=t_*-t_d$ and $e^*=s_*-s_d$.
Using this energy-error characterization, a quadratic energy-error bound for the single-reference CC method \cite{Schneider2009analysis,Rohwedder2013error} and the TNS-TCC method \cite{faulstich2018analysis} follows.

\subsubsection{The Bivariational Approach}
The extended version of the CC method rests on Arponen's bivariational approach \cite{arponen1983variational,lowdin1983stability}.
This unconventional formulation of the CC method parametrizes two independent wave functions and thus makes use of two sets of cluster amplitudes $t=\{ t_\mu \}$ and $\lambda=\{ \lambda_\mu \}$. It gained recent attention in the study~\cite{laestadius2017analysis} and has a major advantage as far as the error analysis is concerned, namely, the energy itself is stationary, i.e., the solution $(t_*,\lambda_*)$ is a critical point of the {\it bivariational energy}, see Eq.~\eqref{Eq:BIVP}. 
Consequently, when the corresponding Galerkin solution $(t_d,\lambda_d)$ is close to the exact solution, a quadratic error estimate is guaranteed.
Subsequently, we elaborate on this further.\newline
\indent Consider the Rayleigh--Ritz quotient, we write
\[
E_0 = \mathcal R(\psi_*)=\min_{\psi\neq 0}\mathcal R(\psi)~.
\]
Hence, $\psi_*$ is a stationary point of $\mathcal R$, i.e., $\mathcal R'(\psi_*)=0$.
By Taylor expanding $\mathcal R$ around $\psi_*$ we obtain the quadratic error estimation for the Rayleigh--Ritz quotient
\[
\vert \mathcal R(\psi)-\mathcal R(\psi_*)\vert 
\leq 
\frac{1}{2}\Vert \mathcal R''(\psi_*)\Vert ~\Vert \psi-\psi_* \Vert^2
+
\mathcal{O}(\Vert \psi-\psi_* \Vert^3)~.
\] 
As mentioned before, the CC formalism does not arise from the Rayleigh--Ritz variational principle. 
However, it can be described by Arponen's bivariational approach, as follows.
Let the bivariate quotient be 
\begin{align}
\label{Eq:BIVP}
\mathcal{B}(\psi, \tilde{\psi})
=
\frac{\langle \tilde{\psi}| \hat H| \psi \rangle}{\langle \tilde{\psi}| \psi \rangle}~.
\end{align}
Eq.~\eqref{Eq:BIVP} can be seen as a generalization of the Rayleigh--Ritz quotient where a stationary point $(\psi_*,\tilde \psi_*)$ is given by a left and right eigenvector of $\hat H$ with corresponding eigenvalue $E=\mathcal{B}(\psi,\tilde \psi)$.
Note that $\mathcal{B}$ is no longer a below bounded functional, hence critical points do not necessarily correspond to extremal points as they do for $\mathcal R$.
In the extended CC theory, the bivariational quotient is studied indirectly by means of the so-called {\it flipped gradient} \cite{laestadius2017analysis}.
Following \cite{arponen1983variational}, we assume $\langle \phi_0|\psi\rangle = \langle \psi|\tilde \psi\rangle =1$ and note that there exists $\hat T$ such that $\psi=e^{\hat T}\phi_0$ (cf. Section~\ref{Sec:ExpAn}). Then $1=\langle \psi|\tilde \psi\rangle = \langle \phi_0| e^{\hat T^\dagger} |\tilde\psi\rangle$ and consequently there exists a cluster operator $\hat \Lambda$ so that $e^{\hat T^\dagger} \tilde\psi = e^{\hat \Lambda} \phi_0$. This defines a smooth coordinate map $\Phi$ from cluster amplitudes $(t,\lambda)$ to wave functions $(\psi,\tilde{\psi})$.
The flipped gradient is then given by $\mathcal F(t,\lambda):=\hat R \,\nabla \mathcal{B}(\Phi(t,\lambda))$, where 
we introduced the flipping map  
\begin{align*}
\hat  R=\begin{pmatrix}
0 &\hat I\\
\hat I &0
\end{pmatrix}.
\end{align*}
Under certain assumptions, $\mathcal F$ is locally strongly monotone~\cite{laestadius2017analysis}. 
By the extended CC approach~\cite{laestadius2017analysis}, $\tilde \psi_*=e^{-\hat T_*^\dagger} e^{\hat \Lambda_*} \phi_0$ and $\psi_*=e^{\hat T_*} \phi_0$ solve the SE if and only if $\mathcal F(t_*,\lambda_*)=0$.
Note that $\mathcal F(t_*,\lambda_*)=0$ implies $\nabla \mathcal B(\Phi(t_*,\lambda_*))=0$ and therewith a quadratic energy error. \newline
\indent Furthermore, by identifying $e^{\hat \Lambda} = \hat I+\hat S$ we obtain from Eq.~\eqref{Eq:BIVP} the CC Lagrangian, i.e.,
\begin{equation}
\begin{aligned}
\label{eq:CCLagrangian}
&\mathcal{B}(e^{\hat T}\phi_0, e^{-\hat T^{\dagger}}e^{\hat \Lambda}\phi_0)
=
\langle \phi_0| e^{-\hat T}\hat He^{\hat T}|\phi_0\rangle\\
&\quad+
\sum_{\mu}s_{\mu}\langle \phi_{\mu}|  e^{-\hat T}\hat He^{\hat T}|\phi_0\rangle =: \mathcal L(t,s) ~.
\end{aligned}
\end{equation}
Introducing the Lagrangian is a general method for optimization with constraints. 
In the special case of CC theory with fixed orbitals, as in this article, Eq.~\eqref{eq:CCLagrangian} demonstrates the equivalence to Arponen's bivariational method \cite{arponen1983variational}. 
In the context of obtaining an efficient evaluation of CC energy gradient, the derivative of the variational functional was obtained by Bartlett \cite{bartlett1986geometrical}.
The functional itself (Eq.~\eqref{eq:CCLagrangian}) was first used in quantum chemistry by Helgaker and J{\o}rgensen \cite{helgaker1988analytical} to derive CC energy derivatives. 
We would also like to mention the related extended CC work of Piecuch and Bartlett~\cite{piecuch1999eomxcc}. Note that their assumption that the reference determinant $\phi_0$ is both a left- and right eigenvector of the doubly similarity transformed $\hat H$ can be rigorously proven in the continuous case (see Lemma~13 in~\cite{laestadius2017analysis}). \newline
\indent Denoting the dual solution $s_* = \{(s_*)_\mu \}$ as in Section~\ref{sec:AN}, it can then be seen that $s_*$ also describes cluster amplitudes parameterizing the wave function $\tilde \psi_*$. 
Indeed, using the relation $e^{\hat \Lambda_*} = \hat I + \hat S_*$, we obtain that $\tilde \psi_* = e^{-\hat T_*^\dagger} (\hat I+\hat S_*) \phi_0$ together with $\psi_* = e^{\hat T_*} \phi_0$ solve the SE corresponding to the same energy $\mathcal B(\psi_*,\tilde\psi_*)$. Assuming non-degeneracy and using the constraint $\langle \tilde \psi_*\vert \psi_* \rangle=1$, we arrive at the condition
\begin{align*}
e^{-\hat T_*^\dagger} (\hat I+\hat S_*) \phi_0 = \frac{1}{\Vert e^{\hat T_*} \phi_0 \Vert^2}e^{\hat T_*} \phi_0
\end{align*}
for the primal and dual solutions $t_*$ and $s_*$.
Thus, from the extended CC theory we have obtained a constraint relating $s_*$ to $t_*$ for the traditional CC method. \newline

\section{Conclusion}
In this article, we have introduced the reader to a local analysis of the CC method and its variations. 
In particular, we have demonstrated that the G\aa rding inequalities for $\hat F$ and $\hat H$ are key as far as a better understanding of the sufficient conditions for a locally unique and quasi-optimal solution of the CC equations is concerned. 
Moreover, these investigations are geared towards an {\it a posteriori} 
criterion of assessing the CC amplitudes from a given computation. This is a mathematical approach that is alternative to the controversial diagnostic suggested in~\cite{Lee1989}.
Indeed, the mathematically derived criteria in Eqs.~\eqref{eq:crit1} and~\eqref{eq:Conj1} use the total $\Vert t \Vert$ and not just the single amplitudes $t_1$. Since the single amplitudes could be removed by an appropriate choice of the reference determinant (i.e., an ideal choice of the basis functions), the sufficient condition for a locally unique solution given by Eq.~\eqref{eq:Conj1} puts constraints on the remaining amplitudes ($t_2,t_3,...$). However, it is not yet a rejection criterion since it only \emph{implies} locally unique and quasi-optimal solutions under certain conditions. 
As outlined, the upper bound in Eq.~\eqref{eq:Conj1} is fundamentally different from previous heuristic and potentially misleading diagnostics~\cite{Lee1989} since the former is derived in a rigorous mathematical framework, where not just the singles amplitudes are taken into consideration.
We have also shown that the condition on the two particle operator in Eq.~\eqref{eq:LipCond} implies a locally unique CC solution. Here, the condition does not explicitly depend on the amplitude norm and might offer a broader understanding of the reliability of a CC solution. Moreover, the derived condition is independent of the chosen single particle operator.
In connection with the extended CC formalism, we have set up a constraint for the exact CC Lagrange multipliers $s_* = \{ (s_*)_\mu \}$, relating them to the exact CC amplitudes $t_*= \{ (t_*)_\mu \}$. Numerical investigations are left for future work.



\bibliography{lib.bib}

\end{document}